\documentclass[pra,twocolumn,showpacs]{revtex4}
\usepackage{float}
\usepackage{makeidx}
\usepackage{amsmath,amssymb,graphics,epsfig,color,times,bbm}

\begin{document}

\title{\large Establishing and storing of deterministic quantum
entanglement among three distant atomic ensembles}
\author{Zhihui Yan$^{1,2\dagger}$}
\author{Liang Wu$^{1\dagger}$}
\author{ Xiaojun Jia$^{1,2}$}
\email{jiaxj@sxu.edu.cn}
\author{Yanhong Liu$^{1}$}
\author{Ruijie Deng$^{1}$}
\author{Shujing Li$^{1,2}$}
\author{Hai Wang$^{1,2}$}
\author{Changde Xie$^{1,2}$}
\author{Kunchi Peng$^{1,2} $}
\affiliation{$^{1}$State Key Laboratory of Quantum Optics and Quantum Optics Devices,
Institute of Opto-Electronics, Shanxi University, Taiyuan, 030006, P. R.
China \\
$^{2}$Collaborative Innovation Center of Extreme Optics, Shanxi University,
Taiyuan 030006, P. R. China \\
}

\begin{abstract}
It is crucial for physical realization
of quantum information networks to first establish entanglement among
multiple space-separated quantum memories and then at a user-controlled
moment to transfer the stored entanglement to quantum channels for
distribution and conveyance of information. Here we present an experimental
demonstration on generation, storage and transfer of deterministic quantum
entanglement among three spatially separated atomic ensembles. The off-line
prepared multipartite entanglement of optical modes is mapped into three
distant atomic ensembles to establish entanglement of atomic spin waves via
electromagnetically-induced-transparency light-matter interaction. Then the
stored atomic entanglement is transferred into a tripartite quadrature
entangled state of light, which is space-separated and can be dynamically
allocated to three quantum channels for conveying quantum information. The
existence of entanglement among released three optical modes verifies that
the system has capacity of preserving multipartite entanglement. The
presented protocol can be directly extended to larger quantum networks with
more nodes.
\end{abstract}

\maketitle

Flying photons or bright optical beams are the best natural quantum
channels, while usually matter systems are employed for memories at quantum
nodes\cite{Kimble1,Polzik0}. Single atoms\cite{Rempe1,Haroche1}, atomic
ensembles\cite{Vuletic1,Lett1,Lam1,Peng2,Yu1,Laurat1}, trapped ions\cite%
{Wineland1,Blatt1,Monroe2}, optomechanics\cite%
{Wang1,Vahala1,Aspelmeyer1,Kiese}, superconductors\cite{Huard1}, solid-state
systems\cite{Gisin1,Tittel1,Sellars1,Imamoglu1} and so on have been applied
as quantum nodes. Especially, atomic ensembles are among the best
candidates for quantum nodes to store and process quantum
information due to the advantage of the collective enhancement of
light-atom interaction\cite{Vuletic1,Lett1,Lam1,Peng2,Yu1,Laurat1}.

The entanglement of discrete quantum variables between two atomic ensembles
has been experimentally achieved by means of Raman scattering approach\cite%
{Kimble3,Pan1} or transferring quantum states of entangled photons into two
atomic systems\cite{Kuzmich1,Kimble4,Guo1}. In 2010, Kimble's group
demonstrated measurement-induced entanglement stored in four atomic memories
and coherent transfer of the atomic entanglement to four photonic channels%
\cite{Kimble2}. For the first time, their experiment proved that a
multipartite entangled W state of atomic ensembles can be transferred into a
photonic mode-entangled W state with the heralded entanglement and thus
showed an advance in the distribution of multipartite entanglement across
quantum networks. Besides above-mentioned schemes based on applying discrete
quantum variables of single photons and atoms, the continuous-variable (CV)
regime provides another avenue toward the realization of quantum
information. To develop CV quantum information networks, CV entanglement
between two macroscopic objects, i.e. atomic ensembles, has been investigated%
\cite{Polzik1,Polzik2}. CV entanglement of spin-wave variances between two
atomic ensembles has been experimentally realized via quantum non-demolition
(QND) measurement\cite{Polzik1} and dissipation mechanism of atomic systems%
\cite{Polzik2}, respectively. For implementing quantum computation\cite%
{Monroe1} and quantum communication\cite{Gisin2}, the established
entanglement has to be stored in atomic memories and then to be
controllably released on demand. Quantum memories for squeezing and
entanglement of light have been theoretically investigated\cite{Yadsan,Tikh}%
, and the storage of CV entanglement between two atomic ensembles
has been experimentally completed\cite{Polzik3}. So far, all
experiments on generation and storage of CV\ entanglement of atomic
systems are concentrated between two
ensembles\cite{Polzik1,Polzik2,Polzik3}. In Ref.[35], a displaced
entangled state of two sideband modes of an optical beam is used for
the initial quantum resource to create entanglement between two
atomic ensembles. To extend this method to multipartite entanglement
more sidebands with different frequency shifts have to be prepared
and the number of entangled sideband modes must be strictly
restricted by the bandwidth of optical parametric amplifier, which
is the device for generating optical entangled state in their
system. On the other hand, it is difficult to space-separate these
entangled optical submodes with small frequency intervals. Although
a narrow-band optical cavity can be utilized for separating these
optical modes, entanglement among them will be significantly
reduced\textbf{\cite{Bachor1}}. These limitations make difficult to
extend the experimental method of Ref.[35] to entangle more
atomic ensembles. Up to now, it still remains a challenge to
entangle more than two remote quantum memories in CV regime.

Here, we present an experimental demonstration on deterministically
establishing, storing and releasing of CV entanglement among three
atomic ensembles. At first, a tripartite optical entangled state is
off-line prepared, and then the entanglement is transferred into
three atomic ensembles located 2.6 meters apart from each other via
electromagnetically-induced-transparency (EIT) interaction. After a
given storage time the preserved atomic entanglement is controllably
released into three separated quantum channels consisting of three
entangled optical submodes. The criterion inequalities and
dependence of entanglement among three released optical submodes on
systematic parameters are theoretically deduced and multipartite
entanglement transfer as well as storage are experimentally proved.
Since the tripartite optical entangled state is generated by
linearly optical transformation of three squeezed states of light,
its three submodes are naturally space-separated\cite{Peng6,Jia1}.
The presented scheme can be directly extended to generate optical
entangled states with more submodes if more squeezed states of light
are available. In this way, entanglement of more atomic ensembles
can be established.

\section*{Results}

\textbf{Experimental Setup.} Fig. 1a describes the experimental
setup for generation, storage and transfer of tripartite entanglement. Three
space-separated submodes $\hat{a}(0)_{\text{S1}}$, $\hat{a}(0)_{\text{S2}}$
and $\hat{a}(0)_{\text{S3}}$ of an optical entangled state off-line prepared
in Part I interact respectively with three atomic memories A$_{1}$, A$_{2}$,
A$_{3}$ located at three distant nodes to generate entanglement of spin
waves among three atomic ensembles. The entanglement is stored within the three
ensembles. Then the preserved entanglement is transferred back into an
optical entangled state with three submodes $\hat{a}(t)_{\text{S1}}$, $\hat{a%
}(t)_{\text{S2}}$ and $\hat{a}(t)_{\text{S3}}$ after a storage time $t$
(Part II). At last, entanglement among three released optical submodes is measured
by three balanced homodyne detectors BHD$_{1-3}$ (Part III).

Three narrow band entangled optical beams tuned to the $\left\vert
5S_{1/2},F=1\right\rangle $ $\leftrightarrow \left\vert 5P_{1/2},F^{\prime
}=1\right\rangle $ transition of Rubidium around 795 nm are
obtained via linearly optical transformation of three squeezed states of
light, which are generated by three degenerate optical parametric amplifiers
(DOPA$_{1-3}$). The DOPA$_{1}$ and DOPA$_{2(3)}$ operating in parametric
amplification and deamplification produce phase and amplitude quadrature
squeezed states $\hat{a}_{\text{S1}}$ and $\hat{a}_{\text{S2(3)}}$,
respectively\cite{Peng1,Peng5}. The squeezing parameters ($r$) for the three
squeezed states are assumed to be identical for simplicity. In fact, three
DOPAs used in our experiment have totally identical configuration and thus
their squeezing parameters are almost the same. The quadrature amplitudes $%
\hat{X}_{\text{S}j}=(\hat{a}_{\text{S}j}+\hat{a}_{\text{S}j}^{+})/\sqrt{2}$
and phases $\hat{P}_{\text{S}j}=(\hat{a}_{\text{S}j}-\hat{a}_{\text{S}%
j}^{+})/\sqrt{2}i$ ($j=1$, $2$, $3$) of output optical beams of
three DOPAs are
expressed as $\hat{X}_{\text{S}1}=$e$^{r}\hat{X}_{\text{S}1}^{(0)}$, $\hat{P}%
_{\text{S}1}=$e$^{-r}\hat{P}_{\text{S}1}^{(0)}$, $\hat{X}_{\text{S}2(3)}=$e$%
^{-r}\hat{X}_{\text{S}2(3)}^{(0)}$ and $\hat{P}_{\text{S}2(3)}=$e$^{r}\hat{P}%
_{\text{S}2(3)}^{(0)}$\cite{Peng5}, where $\hat{X}(\hat{P})_{\text{S}%
1}^{(0)} $, $\hat{X}(\hat{P})_{\text{S}2}^{(0)}$ and $\hat{X}(\hat{P})_{%
\text{S}3}^{(0)}$ are amplitude (phase) quadratures of input optical beams $%
\hat{a}_{\text{S}1}^{(0)}$, $\hat{a}_{\text{S}2}^{(0)}$ and $\hat{a}_{\text{S%
}3}^{(0)}$ for DOPAs, respectively. By interfering three squeezed states of light on BS$%
_{1} $ and BS$_{2}$, we obtain a tripartite optical entangled state with
quantum correlations of both amplitude quadratures $\hat{X}(0)_{\text{L}j}=(%
\hat{a}(0)_{\text{S}j}+\hat{a}(0)_{\text{S}j}^{+})/\sqrt{2}$ and phase
quadratures $\hat{P}(0)_{\text{L}j}=(\hat{a}(0)_{\text{S}j}-\hat{a}(0)_{%
\text{S}j}^{+})/\sqrt{2}i$\cite{Peng5}. Then the three entangled
optical beams are chopped into 500 ns pulses
$\hat{a}(0)_{\text{S}1}$, $\hat{a}(0)_{\text{S}2}$
and $\hat{a}(0)_{\text{S}3}$, with three acoustical-optical modulators AOM$%
_{4-6}$, which are used for the input signals of three atomic ensembles,
i.e. $^{87}$Rb vapour cells A$_{1}$, A$_{2}$ and A$_{3}$.

\begin{figure}[tbp]
\begin{center}
\includegraphics[width=8.6cm]{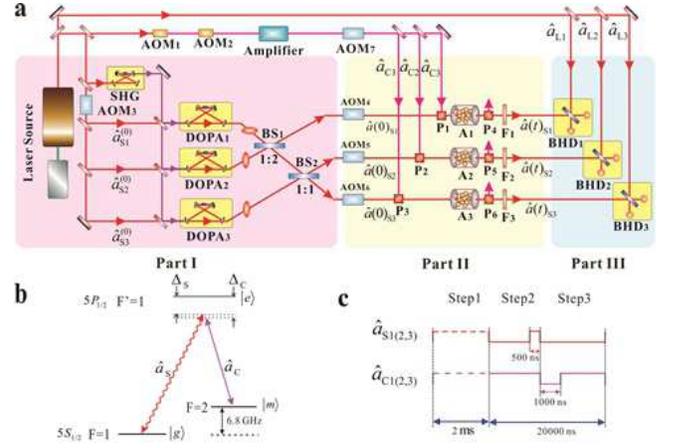}
\end{center}
\par
. .
\caption{\textbf{Schematic diagram.} \textbf{(a)}, Experimental setup. It
includes three parts, Part I is the generation system of tripartite optical
entanglement; Part II expresses the transportation of entanglement of
optical modes to three distant atomic ensembles; Part III is the
entanglement verification system. A$_{1-3}$, atomic ensemble$_{1-3}$; DOPA$%
_{1-3}$, optical parametric amplifier$_{1-3}$; SHG, second harmonic
generator; AOM$_{1-7}$, acousto-optical modulator$_{1-7}$; BS$_{1-2}$, beam
splitter$_{1-2}$; P$_{1-6}$, Glan-Thompson Polarizer$_{1-6}$; F$_{1-3}$,
filter$_{1-3}$; BHD$_{1-3}$, balanced homodyne detector$_{1-3}$; Amplifier,
laser amplifier. \textbf{(b)}, $^{87}$Rb atomic level configuration and
relevant transitions. $\left\vert 5S_{1/2},F=1\right\rangle $ and $%
\left\vert 5S_{1/2},F=2\right\rangle $ play the roles of ground state $%
\left\vert g\right\rangle $ and meta state $\left\vert m\right\rangle $,
respectively, and $\left\vert 5P_{1/2},F^{\prime }=1\right\rangle $ is the
excited state $\left\vert e\right\rangle $. Classical control optical beam $\hat{a}%
_{\text{C}}$ (solid line) and quantum probe optical beam
$\hat{a}_{\text{S}}$ (wavy line) are shown. \textbf{(c)},
Experimental time sequence for control optical beams
$\hat{a}_{\text{C1(2,3)}}$ and signal optical beams
$\hat{a}_{\text{S1(2,3)}}$.}
\end{figure}

\textbf{Quantum state transfer between atomic spin waves and optical
submodes.} The physical mechanism of light-matter interaction used
for the experiment is EIT, which is a transparency phenomenon
induced by optical field in an opaque medium by means of quantum
interference\cite{Marangos1}. M. Fleischhauer and M. D. Lukin have
theoretically demonstrated that when quantum fields propagate in EIT
media there are form-stable quantum excitations associated with such
propagation, named dark-state polaritons, and in this process the
quantum state of light can be ideally transferred to collective
atomic excitations and vice versa. Therefore, EIT effects can be
applied to generate nonclassical states of atomic ensembles, to
store optical quantum states, and reversibly to release stored
quantum states into optical channels,
respectively\cite{Lukin2,Giacobino1,Lvovsky1,Furusawa1}. An atomic
ensemble is represented by total angular momentum operator of
collective atomic spins $\hat{J}=\sum_{i}\left\vert g\right\rangle
\left\langle m\right\vert $, and y, z-components of the collective
atomic angular
momentum play the role of canonical variables, i.e. $\hat{X}_{\text{A}}=(\hat{J%
}+\hat{J}^{+})/\sqrt{2}=\hat{J}_{y}/\sqrt{\langle \hat{J}_{x}\rangle }$, $%
\hat{P}_{\text{A}}=(\hat{J}-\hat{J}^{+})/\sqrt{2}i=\hat{J}_{z}/\sqrt{\langle
\hat{J}_{x}\rangle }$\cite{Polzik3}. When control optical beams are
adiabatically switched off, the mapping relations of amplitude
(phase) quadratures from input optical submodes
$\hat{X}(\hat{P})(0)_{\text{L}j}$ to atomic spin waves
$\hat{X}(\hat{P})(t)_{\text{A}j}$ after a storage time $t$ are
expressed by\cite{Ou1}:

\begin{eqnarray}
\hat{X}(t)_{\text{A}j} &=&\sqrt{\eta _{_{\text{M}}}}\hat{X}(0)_{\text{L}j}+%
\sqrt{1-\eta _{_{\text{M}}}}\hat{X}_{\text{A}j}^{\text{vac}}\text{,} \\
\hat{P}(t)_{\text{A}j} &=&\sqrt{\eta _{_{\text{M}}}}\hat{P}(0)_{\text{L}j}+%
\sqrt{1-\eta _{_{\text{M}}}}\hat{P}_{\text{A}j}^{\text{vac}}\text{,}  \notag
\end{eqnarray}%
where the mapping efficiency from input optical submodes to atomic spin
waves is $\eta _{_{\text{M}}}=\eta _{_{\text{T}}}\eta _{_{\text{W}}}$e$%
^{-t/\tau _{s}}$, $\eta _{_{\text{T}}}$ is the optical transmission
efficiency, $\eta _{_{\text{W}}}$ is the storage efficiency of light in
atomic ensemble, and $\tau _{\text{s}}$ is the storage lifetime limited by
atomic decoherence. The vacuum quadrature noises of atomic ensembles $\hat{X}%
(\hat{P})_{\text{A}j}^{\text{vac}}$ are introduced by limited mapping
efficiency $\eta _{_{\text{M}}}$. Since canonical quadrature operators of
atomic spin waves obey the same commutation relation with that of Gaussian
optical states, i.e. $[\hat{X}_{\text{A}},\hat{P}_{\text{A}}]=i$, using
similar procedure of deducing full tripartite inseparability criteria
provided by Loock \textit{et al.}, we can obtain a set of analogous
criterion inequalities for atomic spin waves (see also the Method section).

The stored atomic entanglement can be transferred to tripartite entanglement
among output three optical submodes $\hat{a}(t)_{\text{P}1}$, $\hat{a}(t)_{%
\text{P}2}$ and $\hat{a}(t)_{\text{P}3}$ by turning on control
optical beams. The
quadrature amplitudes and phases of released submodes, $\hat{X}(t)_{\text{L}%
j}=(\hat{a}(t)_{\text{P}j}+\hat{a}(t)_{\text{P}j}^{+})/\sqrt{2}$ and $\hat{P}%
(t)_{\text{L}j}=(\hat{a}(t)_{\text{P}j}-\hat{a}(t)_{\text{P}j}^{+})/\sqrt{2}%
i $ in terms of quadratures for atomic spin waves $\hat{X}(\hat{P})(t)_{%
\text{A}j}$ after a storage time $t$ are expressed by\cite{Ou1}:

\begin{eqnarray}
\hat{X}(t)_{\text{L}j} &=&-\sqrt{\eta _{_{\text{M}}}^{\prime }}\hat{X}(t)_{%
\text{A}j}+\sqrt{1-\eta _{_{\text{M}}}^{\prime }}\hat{X}_{\text{L}j}^{\text{%
vac}}\text{,} \\
\hat{P}(t)_{\text{L}j} &=&-\sqrt{\eta _{_{\text{M}}}^{\prime }}\hat{P}(t)_{%
\text{A}j}+\sqrt{1-\eta _{_{\text{M}}}^{\prime }}\hat{P}_{\text{L}j}^{\text{%
vac}}\text{,}  \notag
\end{eqnarray}%
where the mapping efficiency from atomic spin waves to optical submodes $%
\eta _{_{\text{M}}}^{\prime }$ is the retrieval efficiency from atomic
ensembles to light. The vacuum quadrature noises of optical submodes $\hat{X}%
(\hat{P})_{\text{L}j}^{\text{vac}}$ are introduced by the read process.

The full tripartite inseparability criteria for released optical
modes are given by\cite{Loock1}:

\begin{eqnarray}
I(t)_{\text{L}1} &=&\langle \delta ^{2}(\hat{X}(t)_{\text{L}2}-\hat{X}(t)_{%
\text{L}3})\rangle /2+\langle \delta ^{2}(g_{\text{L}1}^{\prime }\hat{P}(t)_{%
\text{L}1}  \notag \\
&&+\hat{P}(t)_{\text{L}2}+\hat{P}(t)_{\text{L}3})\rangle /2\geqslant 1\text{,%
} \\
I(t)_{\text{L}2} &=&\langle \delta ^{2}(\hat{X}(t)_{\text{L}1}-\hat{X}(t)_{%
\text{L}3})\rangle /2+\langle \delta ^{2}(\hat{P}(t)_{\text{L}1}  \notag \\
&&+g_{\text{L}2}^{\prime }\hat{P}(t)_{\text{L}2}+\hat{P}(t)_{\text{L}%
3})\rangle /2\geqslant 1\text{,}  \notag \\
I(t)_{\text{L}3} &=&\langle \delta ^{2}(\hat{X}(t)_{\text{L}1}-\hat{X}(t)_{%
\text{L}2})\rangle /2+\langle \delta ^{2}(\hat{P}(t)_{\text{L}1}  \notag \\
&&+\hat{P}(t)_{\text{L}2}+g_{\text{L}3}^{\prime }\hat{P}(t)_{\text{L}%
3})\rangle /2\geqslant 1\text{.}  \notag
\end{eqnarray}%
If any two in the three inequalities are simultaneously violated, the three
submodes form a tripartite Greenberger-Horne-Zeilinger-like (GHZ-like)
entangled state of light, where $g_{\text{L1}}^{\prime }$, $g_{\text{L2}%
}^{\prime }$ and $g_{\text{L3}}^{\prime }$ are the gain factors for
optimizing the correlation variances for released submodes. From Eqs. (1)
and (2), it can be seen that entanglement is limited by total mapping
efficiencies $\eta $ ($\eta =\eta _{_{\text{M}}}\eta _{_{\text{M}}}^{\prime
} $) as well as squeezing parameter $r$. When the squeezing parameters $r$
for three DOPAs and the total mapping efficiencies $\eta $ for three atomic
memories are the same, the values of left sides of three inequalities in Eq.
(3) are identical, $I(t)_{\text{L1}}=I(t)_{\text{L2}}=I(t)_{\text{L3}}=I(t)_{%
\text{L}}$. The smaller $I(t)_{\text{L}}$ is, the higher the
entanglement is. Fig. 2 shows the dependence of correlation variance
combinations for three released submodes on the squeezing parameter $r$ of
initial squeezed states and the total mapping efficiency $\eta $ with the
storage time of 1000 ns. We can see that the combinations are reduced with
the increase of squeezing parameter $r$ and total mapping efficiency $\eta $
(see also the Method section).

\begin{figure}[tbp]
\begin{center}
\includegraphics[width=8.6cm]{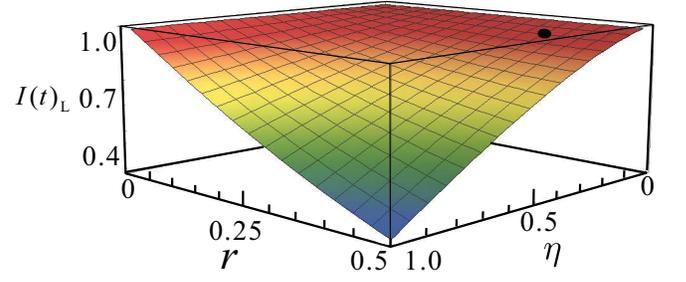}
\end{center}
\caption{The dependence of combinations of normalized quantum correlation
variances among three released submodes after a storage time of 1 $\protect%
\mu s$ on the squeezing parameter $r$ of three DOPAs and the total mapping
efficiency $\protect\eta $, where the gains $g_{\text{L1-L3}}^{\prime }$ are
taken as the optimal gain $g_{\text{L1-L3}}^{\prime \text{opt}}$. \textbf{%
The dot corresponds the experimental result of $I(t)_{\text{L}}=0.96\pm 0.01$%
, where the squeezing parameter $r$ is $0.38$ and total mapping efficiency $%
\protect\eta $ is about $16\%$.}}
\end{figure}

When the control optical beams $\hat{a}_{\text{C1}}$,
$\hat{a}_{\text{C2}}$ and $\hat{a}_{\text{C3}}$, which are tuned to
$\left\vert 5S_{1/2},F=2\right\rangle $ $\leftrightarrow \left\vert
5P_{1/2},F^{\prime }=1\right\rangle $ transition of
Rubidium, are adiabatically
switched off, optical entanglement among three input submodes $\hat{a}(0)_{%
\text{S1}}$, $\hat{a}(0)_{\text{S2}}$ and $\hat{a}(0)_{\text{S3}}$ is
transferred to atomic entanglement of spin waves according to Eq. (1) via
EIT interaction. After a storage time of 1 $\mu s$, control optical beams $\hat{a}_{%
\text{C1}}$, $\hat{a}_{\text{C2}}$ and $\hat{a}_{\text{C3}}$ are switched on
again, three optical submodes $\hat{a}(t)_{\text{S1}}$, $\hat{a}(t)_{\text{S2%
}}$ and $\hat{a}(t)_{\text{S3}}$ are released. The combinations of
correlation variances in Eq. (3) are measured with three time domain
BHD$_{1-3}$. The intensive coherent light $\hat{a}_{\text{L1}}$,
$\hat{a}_{\text{L2}}$ and $\hat{a}_{\text{L3}}$ are utilized as
local oscillators of BHD$_{1-3}$. The control and signal optical
beams with orthogonally linear polarizations are combined on
Glan-Thompson Polarizers (P$_{1-3}$) before atomic cells, and
control optical beams are filtered out from the signal optical beams
by Glan-Thompson Polarizers (P$_{4-6}$) and etalon filters
(F$_{1-3}$). In storage and retrieval procedures 10000 traces of BHD
output signals with 20 G samples/s are digitally filtered with a
bandpass filter of 2.5 MHz and averaged to obtain the optimal
entangled degree. In this case the low-frequency sideband noises
resulting from pumping laser of DOPAs and atomic ensembles, as well as high
frequency noises coming from parametric conversion in DOPAs and EIT process in atomic ensembles have been are filtered out.
When both control and signal optical beams are blocked and only the
local oscillators are remained, outputs of BHDs stand for the corresponding vacuum noise level%
\cite{Ou0}.

\begin{widetext}
\begin{table}[h]
\caption{The values of normalized correlation variances for different
combinations.}\centering
\begin{tabular}{cccc}
Correlation variances & ~~Values for input~~ & ~~Values for atomic ~~ &
~~Values for released ~~ \\
for different combinations & submodes (dB) & spin waves (dB) & submodes (dB)
\\ \hline
$\langle \delta ^{2}(\hat{X}_{2}-\hat{X}_{3})\rangle $ & -3.30$\pm $0.05 &
-0.56$\pm $0.03 & -0.37$\pm $0.03 \\
$\langle \delta ^{2}(g_{1}\hat{P}_{1}+\hat{P}_{2}+\hat{P}_{3})\rangle $ &
-2.93$\pm $0.05 & -0.15$\pm $0.02 & -0.10$\pm $0.02 \\
$\langle \delta ^{2}(\hat{X}_{1}-\hat{X}_{3})\rangle $ & -3.25$\pm $0.05 &
-0.53$\pm $0.03 & -0.35$\pm $0.03 \\
$\langle \delta ^{2}(\hat{P}_{1}+g_{2}\hat{P}_{2}+\hat{P}_{3})\rangle $ &
-2.91$\pm $0.05 & -0.15$\pm $0.02 & -0.10$\pm $0.02 \\
$\langle \delta ^{2}(\hat{X}_{1}-\hat{X}_{2})\rangle $ & -3.25$\pm $0.05 &
-0.52$\pm $0.03 & -0.34$\pm $0.03 \\
$\langle \delta ^{2}(\hat{P}_{1}+\hat{P}_{2}+g_{3}\hat{P}_{3})\rangle $ &
-2.90$\pm $0.05 & -0.14$\pm $0.02 & -0.09$\pm $0.02 \\ \hline
\end{tabular}%
\end{table}
\end{widetext}

\begin{figure}[tbp]
\begin{center}
\includegraphics[width=8.6cm]{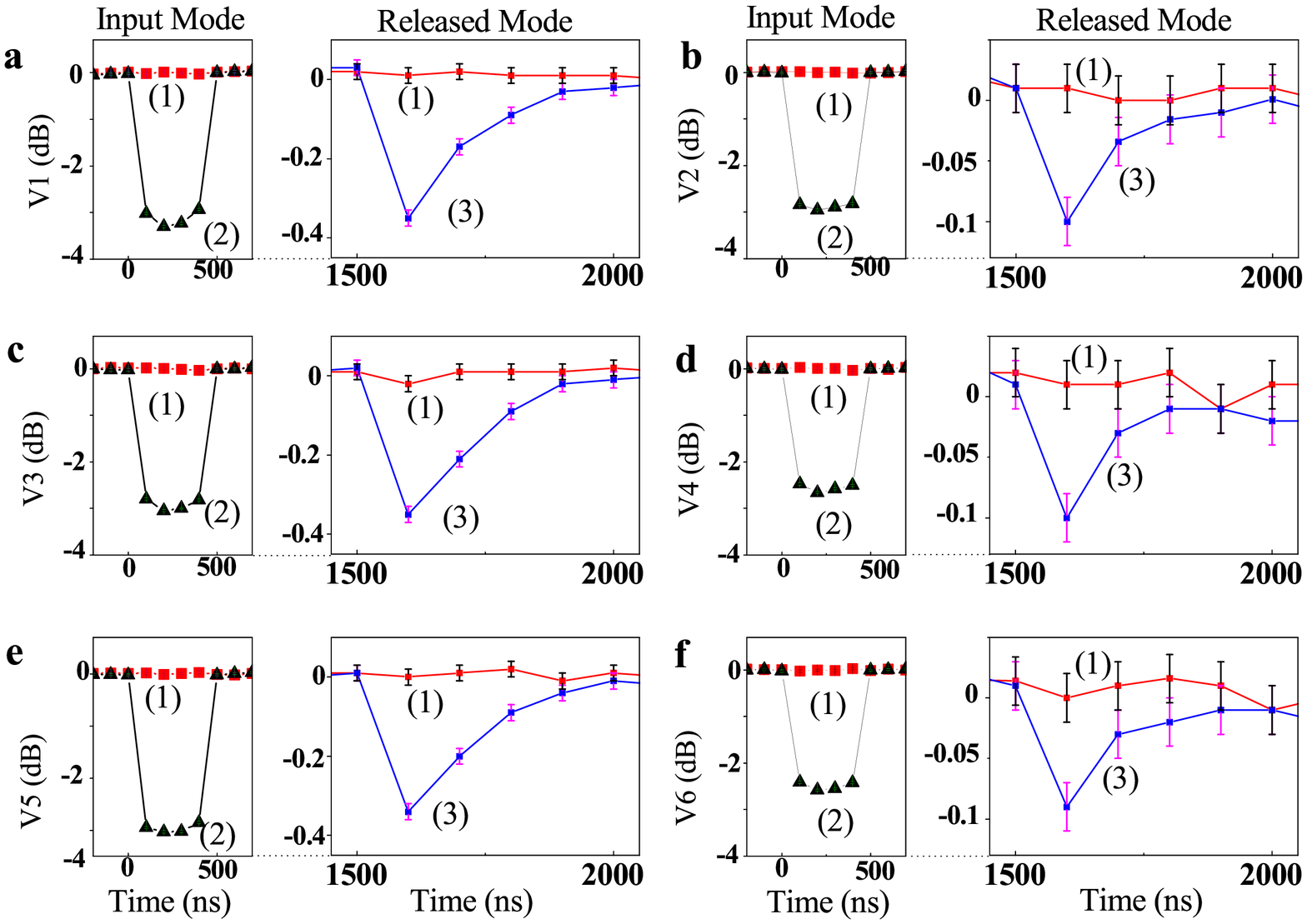}
\end{center}
\caption{Measured normalized correlation variances of input and released
optical submodes. Trace (1) is the vacuum noise level. Trace (2) is the
correlation variances of the original input optical submodes. \textbf{(a)}%
,V1 ($\langle \protect\delta ^{2}(\hat{X}(0)_{\text{L2}}-\hat{X}(0)_{\text{L3%
}})\rangle $), \textbf{(b)}, V2 ($\langle \protect\delta ^{2}(g_{\text{L1}}^{%
\text{opt}}\hat{P}(0)_{\text{L1}}+\hat{P}(0)_{\text{L2}}+\hat{P}(0)_{\text{L3%
}})\rangle $), \textbf{(c)}, V3 ($\langle \protect\delta ^{2}(\hat{X}(0)_{%
\text{L1}}-\hat{X}(0)_{\text{L3}})\rangle $), \textbf{(d)}, V4 ($\langle
\protect\delta ^{2}(\hat{P}(0)_{\text{L1}}+g_{\text{L2}}^{\text{opt}}\hat{P}%
(0)_{\text{L2}}+\hat{P}(0)_{\text{L3}})\rangle $), \textbf{(e)}, V5 ($%
\langle \protect\delta ^{2}(\hat{X}(0)_{\text{L1}}-\hat{X}(0)_{\text{L2}%
})\rangle $), \textbf{(f)}, V6 ($\langle \protect\delta ^{2}(\hat{P}(0)_{%
\text{L1}}+\hat{P}(0)_{\text{L2}}+g_{\text{L3}}^{\text{opt}}\hat{P}(0)_{%
\text{L3}})\rangle $). Trace (3) is corresponding correlation variances of
released optical submodes after a storage time of 1000 ns. Error bars
represent $\pm $ one standard error and are obtained with the statistics of
the measured correlation variances.}
\end{figure}

\textbf{Experimental Results.} The normalized correlation variances for
different combinations of quadrature components are given in Table. 1, where
gain factors ($g_{1}$, $g_{2}$ and $g_{3}$) are chosen as the optimal gains
for minimizing the corresponding correlation variances. The correlation
variances of input and released submodes are directly measured with three
sets of BHDs. The normalized correlation variances among three atomic
ensembles are inferred from Eq. (2), where the mapping efficiency $\eta _{_{%
\text{M}}}^{\prime }$ is about $68\%$ for our experimental system. The
measured normalized quantum correlation variances are shown in Fig. 3. The
squeezing parameter $r$ is $0.38$ and total mapping efficiency $\eta $ is
about $16\%$. The combination of correlation variances for three released
submodes is $I(t)_{\text{L}}=0.96\pm 0.01$, which is less than 1, thus
according to criterion inequalities Eq. (3) the entanglement among released
submodes is verified. The value is in agreement with the theoretically
calculated result, which is marked with a black dot in Fig. 2.
Because of the limitation of total mapping efficiency, the correlation
variances of released entangled state are much higher than that of input
state. However, correlation variances below the corresponding vacuum noise
level in Fig. 3 and the violation of criterion inequalities in Eq. (3)
certainly prove the existence of tripartite entanglement among three optical
submodes released from atomic spin waves of three atomic ensembles. Thus the
tripartite GHZ-like entanglement among three atomic ensembles is
experimentally demonstrated.

\section*{Discussion}

In summary, deterministic quantum entanglement among three spatially
separated quantum nodes is experimentally generated, stored and transferred.
Within storage lifetime the multipartite entanglement is preserved in three
space-separated atomic ensembles, and then at a desirable time the stored
atomic entanglement can be controllably converted into three optical
submodes to be quantum channels. Our work shows that multipartite CV
entanglement can be established among remote macroscopic objects by
transferring off-line prepared entanglement of optical beams into atomic
ensembles via EIT interaction. Since unconditional CV entanglement among
multipartite optical modes has been experimentally accomplished\cite%
{Peng6,Furusawa2,Treps1}, mature quantum optical technology can be used for
realizing entanglement of more nodes in a quantum networks. The obtained
entanglement among atomic ensembles depends on mapping efficiency $\eta _{_{%
\text{M}}}$ from light to atoms and initial squeezing parameter $r$. The
higher the squeezing and the mapping efficiency are, the better the
entanglement among atomic ensembles is. In the presented experimental
system, the total mapping efficiency is mainly limited by the optical
transmission loss and memory (storage and retrieval) efficiency. The
transmission loss (about 18\%) comes from the optical losses of atomic
cells, etalon filters, Glan-Thompson Polarizers and other optical
components, which can be further reduced if better optical elements with
lower losses are available.

The increased excess noises in released submodes from memories
originate from fluorescence and coherent emission as well as spurious
fluctuations in signal channels induced by the control optical beam\cite%
{Lvovsky1}. On the other hand, since both EIT and four-wave mixing (FWM)
effects are simultaneously generated in an ensemble of hot atoms\cite%
{Phillips1}, the control optical beam acts as a far-detuned field on
the signal transition in the undesired FWM process and spontaneously
generates an \textquotedblleft idler\textquotedblright\ field, which
also can form excess noises. These mechanisms resulting in excess
noises always exist in EIT light-matter interaction for any
input states, whatever vacuum state, squeezed vacuum state,
entangled state or others. Therefore some schemes
used for improving EIT memory efficiency of classical signals\cite%
{Lobino1,Phillips1,Lauk1}, such as decreasing detuning of probe and
control optical beams, increasing the power of control optical beam
and enhancing the temperature of atomic vapour, will unavoidably
introduce more excess noises into atomic media and reduce quantum
correlations among atomic ensembles. However, by
optimizing experimental parameters, excess noises can be
minimized for a given EIT experimental system\cite%
{Lvovsky1,Giacobino1,Phillips1}. 

It has been demonstrated that mapping efficiency can be significantly
improved by means of the technology of optical cavity enhancement without
introducing excess noise\cite{Pan3,Walmsley1} and the storage lifetime can
be dramatically increased if thermal atomic ensembles are replaced by cold
atoms confined in three dimensional optical lattice\cite{Pan3}. The
generation systems of optical squeezed states with high squeezing up to 15
dB has been available today\cite{Schnabel1}, which can provide initial high
quality quantum resource for establishing better multipartite entanglement
among atomic memories. The presented scheme opens up a new possibility for
constructing future quantum internet\cite{Kimble1} and implementing
distribution quantum computation based on the use of deterministic CV
entanglement resources of light and atomic memories\cite{Stace1}.

\section*{Methods}

\textbf{Experimental realization of tripartite entangled state of
light.} A Ti:sapphire laser (Coherent MBR-110) pumped by green laser
(Yuguang DPSS FG-VIIIB) outputs 3 W laser at about 795 nm
wavelength, which is used for pumping light of second harmonic
generator (SHG) and seed lights of DOPAs. The configuration of
optical cavities for SHG and three DOPAs is identical bow tie type ring
cavity with a $1\times 2\times 10$ mm periodically poled KTiOPO4
(PPKTP) crystal. Three DOPAs are pumped by the second harmonic
fields at about 398 nm from SHG and the fundamental waves from SHG
are utilized
as three injected seed fields ($\hat{a}_{\text{S1}}^{(0)}$, $\hat{a}_{\text{%
S2}}^{(0)}$ and $\hat{a}_{\text{S3}}^{(0)}$). DOPA$_{1}$ and
DOPA$_{2(3)}$ are operated at parametric amplification and
deamplification to produce quadrature phase and amplitude squeezed
state of light, respectively\cite{Peng2}.

The three squeezed optical beams are interfered on two optical beam
splitters. The quadrature phase squeezed field from DOPA$_{1}$ ($\hat{a}_{%
\text{S1}}$) and the quadrature amplitude squeezed field from DOPA$_{2}$ ($%
\hat{a}_{\text{S2}}$) are firstly interfered on a beam splitter
(BS1) with the ratio of R:T=1:2 (R: reflectivity and T:
transmissivity). Then, one of two
output optical beams from BS1 and the quadrature amplitude squeezed light from DOPA$%
_{3}$ ($\hat{a}_{\text{S3}}$) are interfered on BS2 with the ratio
of R:T=1:1. The relative phase between the two input optical beams
on BS1(2) is kept at $2k\pi $ ($k$ is integer). Finally three
entangled optical beams are chopped into three optical pulses $\hat{a}(0)_{\text{S1}}$, $\hat{a}(0)_{%
\text{S2}}$ and $\hat{a}(0)_{\text{S3}}$ by three AOMs.
The three optical pulses are respectively injected into three atomic
ensembles to be the input optical submodes. The
quadratures amplitudes $\hat{X}(0)_{\text{L}j}=(\hat{a}(0)_{\text{S}j}+\hat{a%
}(0)_{\text{S}j}^{+})/\sqrt{2}$ and phases $\hat{P}(0)_{\text{L}j}=(\hat{a}%
(0)_{\text{S}j}-\hat{a}(0)_{\text{S}j}^{+})/\sqrt{2}i$ ($j=1,2,3$)
of input optical submodes are expressed by\cite{Furusawa2,Peng2}:

\begin{equation}
\hat{X}(0)_{\text{L}1}=\sqrt{\frac{1}{3}}\text{e}^{r}\hat{X}_{\text{S}%
1}^{(0)}+\sqrt{\frac{2}{3}}\text{e}^{-r}\hat{X}_{\text{S}2}^{(0)}\text{,}
\end{equation}%
\begin{eqnarray*}
\hat{P}(0)_{\text{L}1} &=&\sqrt{\frac{1}{3}}\text{e}^{-r}\hat{P}_{\text{S}%
1}^{(0)}+\sqrt{\frac{2}{3}}\text{e}^{r}\hat{P}_{\text{S}2}^{(0)}\text{,} \\
\hat{X}(0)_{\text{L}2} &=&\sqrt{\frac{1}{3}}\text{e}^{r}\hat{X}_{\text{S}%
1}^{(0)}-\sqrt{\frac{1}{6}}\text{e}^{-r}\hat{X}_{\text{S}2}^{(0)}+\sqrt{%
\frac{1}{2}}\text{e}^{-r}\hat{X}_{\text{S}3}^{(0)}\text{,} \\
\hat{P}(0)_{\text{L}2} &=&\sqrt{\frac{1}{3}}\text{e}^{-r}\hat{P}_{\text{S}%
1}^{(0)}-\sqrt{\frac{1}{6}}\text{e}^{r}\hat{P}_{\text{S}2}^{(0)}+\sqrt{\frac{%
1}{2}}\text{e}^{r}\hat{P}_{\text{S}3}^{(0)}\text{,} \\
\hat{X}(0)_{\text{L}3} &=&\sqrt{\frac{1}{3}}\text{e}^{r}\hat{X}_{\text{S}%
1}^{(0)}-\sqrt{\frac{1}{6}}\text{e}^{-r}\hat{X}_{\text{S}2}^{(0)}-\sqrt{%
\frac{1}{2}}\text{e}^{-r}\hat{X}_{\text{S}3}^{(0)}\text{,} \\
\hat{P}(0)_{\text{L}3} &=&\sqrt{\frac{1}{3}}\text{e}^{-r}\hat{P}_{\text{S}%
1}^{(0)}-\sqrt{\frac{1}{6}}\text{e}^{r}\hat{P}_{\text{S}2}^{(0)}-\sqrt{\frac{%
1}{2}}\text{e}^{r}\hat{P}_{\text{S}3}^{(0)}\text{,}
\end{eqnarray*}%
respectively.

The inequalities of full inseparability criteria for input tripartite
entangled states of light are\cite{Loock1}:

\begin{widetext}
\begin{eqnarray}
I(0)_{\text{L}1} &=&\langle \delta ^{2}(\hat{X}(0)_{\text{L}2}-\hat{X}(0)_{%
\text{L}3})\rangle /2+\langle \delta ^{2}(g_{\text{L}1}\hat{P}(0)_{\text{L}%
1}+\hat{P}(0)_{\text{L}2}+\hat{P}(0)_{\text{L}3})\rangle /2\geqslant 1\text{,%
} \\
I(0)_{\text{L}2} &=&\langle \delta ^{2}(\hat{X}(0)_{\text{L}1}-\hat{X}(0)_{%
\text{L}3})\rangle /2+\langle \delta ^{2}(\hat{P}(0)_{\text{L}1}+g_{\text{L}%
2}\hat{P}(0)_{\text{L}2}+\hat{P}(0)_{\text{L}3})\rangle /2\geqslant 1\text{,}
\notag \\
I(0)_{\text{L}3} &=&\langle \delta ^{2}(\hat{X}(0)_{\text{L}1}-\hat{X}(0)_{%
\text{L}2})\rangle /2+\langle \delta ^{2}(\hat{P}(0)_{\text{L}1}+\hat{P}(0)_{%
\text{L}2}+g_{\text{L}3}\hat{P}(0)_{\text{L}3})\rangle /2\geqslant 1\text{.}
\notag
\end{eqnarray}
\end{widetext}

If any two in the three inequalities are simultaneously violated, the three
submodes form a tripartite GHZ-like entangled state, where $g_{\text{L}%
1}$, $g_{\text{L}2}$ and $g_{\text{L}3}$ are gain factors for
minimizing correlation variances of the input tripartite entangled
state of light.

When the squeezing parameter $r$ ($r\geqslant 0$) for three DOPAs is the
same, the gain factors in Eq. (5) should be the same, i.e. $g_{\text{L}1}=g_{%
\text{L}2}=g_{\text{L}3}=g_{\text{L}}$, and the values of left sides of
three inequalities are identical, i.e. $I(0)_{\text{L}1}=I(0)_{\text{L}2}=I(0)_{\text{L}3}=I(0)_{%
\text{L}}$. Using Eqs. (4) and (5), the combination of
normalized quantum correlation variances for input optical beams is
obtained:

\begin{equation}
I(0)_{\text{L}}=\frac{12\text{e}^{-2r}+2(g_{\text{L}}+2)^{2}\text{e}%
^{-2r}+4(g_{\text{L}}-1)^{2}\text{e}^{2r}}{24}\text{.}
\end{equation}%
Calculating the minimum value of Eq. (6), we get the optimal gain factor $g_{%
\text{L}}^{\text{opt}}$:

\begin{equation}
g_{\text{L}}^{\text{opt}}=\frac{2\text{e}^{4r}-2}{2\text{e}^{4r}+1}\text{.}
\end{equation}%
If squeezing parameter $r$ is larger than 0, the combination of correlation
variances will be less than 1 with the optimal gain factor and the input
optical submodes of atomic ensembles are in a tripartite GHZ-like entangled
state.

\textbf{Establishing and storing of tripartite atomic-ensemble entanglement.}
In EIT memory medium, quantum state can be mapped from input optical submode
$\hat{a}_{\text{S}}$ into atomic spin wave $\hat{J}$ and vice versus under
the interaction with a the strong control optical beam $\hat{a}_{\text{C}}$\cite%
{Fleischhauer1,Lukin2}. The control optical beam is treated as
classical optical beam $A_{\text{C}}$ because it is much more
intensive than the signal
optical modes. In EIT process, the effective interaction Hamiltonian $\hat{H}_{\text{%
EIT}}$ between signal optical mode $\hat{a}_{\text{S}}$ and atomic
spin wave $\hat{J}$ is given by\cite{Polzik1,Ou1}:

\begin{equation}
\hat{H}_{\text{EIT}}=i\hbar \kappa A_{\text{C}}\hat{a}_{\text{P}}\hat{J}%
^{+}-i\hbar \kappa A_{\text{C}}\hat{a}_{\text{P}}^{+}\hat{J}\text{,}
\end{equation}%
which is similar to a beam-splitter interaction, where $\kappa $ stands for
the interaction constant between light and atoms.

By solving Heisenberg motion equations with the Hamiltonian $\hat{H}_{%
\text{EIT}}$ (Eq. (8)), we obtain the expressions of quantum storage process
and the mapping relations (Eq. (1)) of amplitude and phase quadratures from
input optical submodes $\hat{X}(\hat{P})(0)_{\text{L}j}$ to atomic spin
waves $\hat{X}(\hat{P})(t)_{\text{A}j}$ after a storage time $t$.

When the control optical beams are turned on, the input signal
submodes are compressed in atomic ensembles due to the slow
propagation under EIT interaction. On the moment of simultaneously
shutting off three control optical beams, quantum entanglement among
three pulse submodes will be mapped into atomic spin waves in the
three ensembles, where the control optical beam plays the role of
writing process. Using Eqs. (1) and (4), the
amplitude (phase) quadratures of atomic spin waves $\hat{X}(\hat{P})(t)_{%
\text{A}j}$ after a storage time $t$ are obtained:

\begin{widetext}
\begin{eqnarray}
\hat{X}(t)_{\text{A}1} &=&\sqrt{\frac{\eta _{_{\text{M}}}}{3}}\text{e}^{r}%
\hat{X}_{\text{S}1}^{(0)}+\sqrt{\frac{2\eta _{_{\text{M}}}}{3}}\text{e}^{-r}%
\hat{X}_{\text{S}2}^{(0)}+\sqrt{1-\eta _{_{\text{M}}}}\hat{X}_{\text{A}1}^{%
\text{vac}}\text{,} \\
\hat{P}(t)_{\text{A}1} &=&\sqrt{\frac{\eta _{_{\text{M}}}}{3}}\text{e}^{-r}%
\hat{P}_{\text{S}1}^{(0)}+\sqrt{\frac{2\eta _{_{\text{M}}}}{3}}\text{e}^{r}%
\hat{P}_{\text{S}2}^{(0)}+\sqrt{1-\eta _{_{\text{M}}}}\hat{P}_{\text{A}1}^{%
\text{vac}}\text{,}  \notag \\
\hat{X}(t)_{\text{A}2} &=&\sqrt{\frac{\eta _{_{\text{M}}}}{3}}\text{e}^{r}%
\hat{X}_{\text{S}1}^{(0)}-\sqrt{\frac{\eta _{_{\text{M}}}}{6}}\text{e}^{-r}%
\hat{X}_{\text{S}2}^{(0)}+\sqrt{\frac{\eta _{_{\text{M}}}}{2}}\text{e}^{-r}%
\hat{X}_{\text{S}3}^{(0)}+\sqrt{1-\eta _{_{\text{M}}}}\hat{X}_{\text{A}2}^{%
\text{vac}}\text{,}  \notag \\
\hat{P}(t)_{\text{A}2} &=&\sqrt{\frac{\eta _{_{\text{M}}}}{3}}\text{e}^{-r}%
\hat{P}_{\text{S}1}^{(0)}-\sqrt{\frac{\eta _{_{\text{M}}}}{6}}\text{e}^{r}%
\hat{P}_{\text{S}2}^{(0)}+\sqrt{\frac{\eta _{_{\text{M}}}}{2}}\text{e}^{r}%
\hat{P}_{\text{S}3}^{(0)}+\sqrt{1-\eta _{_{\text{M}}}}\hat{P}_{\text{A}2}^{%
\text{vac}}\text{,}  \notag \\
\hat{X}(t)_{\text{A}3} &=&\sqrt{\frac{\eta _{_{\text{M}}}}{3}}\text{e}^{r}%
\hat{X}_{\text{S}1}^{(0)}-\sqrt{\frac{\eta _{_{\text{M}}}}{6}}\text{e}^{-r}%
\hat{X}_{\text{S}2}^{(0)}-\sqrt{\frac{\eta _{_{\text{M}}}}{2}}\text{e}^{-r}%
\hat{X}_{\text{S}3}^{(0)}+\sqrt{1-\eta _{_{\text{M}}}}\hat{X}_{\text{A}3}^{%
\text{vac}}\text{,}  \notag \\
\hat{P}(t)_{\text{A}3} &=&\sqrt{\frac{\eta _{_{\text{M}}}}{3}}\text{e}^{-r}%
\hat{P}_{\text{S}1}^{(0)}-\sqrt{\frac{\eta _{_{\text{M}}}}{6}}\text{e}^{r}%
\hat{P}_{\text{S}2}^{(0)}-\sqrt{\frac{\eta _{_{\text{M}}}}{2}}\text{e}^{r}%
\hat{P}_{\text{S}3}^{(0)}+\sqrt{1-\eta _{_{\text{M}}}}\hat{P}_{\text{A}3}^{%
\text{vac}}\text{,}  \notag
\end{eqnarray}
\end{widetext}

respectively.

Since canonical quadrature operators of atomic spin waves obey the same
commutation relation with that of Gaussian optical states, i.e. $[\hat{X}_{%
\text{A}},\hat{P}_{\text{A}}]=i$, using similar procedure of
deducing full tripartite inseparability criteria provided by Loock
\textit{et al.}\cite{Loock1}, we obtain a set of analogous criterion
inequalities for atomic spin waves:

\begin{widetext}
\begin{eqnarray}
I(t)_{\text{A1}} &=&\langle \delta ^{2}(\hat{X}(t)_{\text{A2}}-\hat{X}(t)_{%
\text{A3}})\rangle /2+\langle \delta ^{2}(g_{\text{A1}}\hat{P}(t)_{\text{A1}%
}+\hat{P}(t)_{\text{A2}}+\hat{P}(t)_{\text{A3}})\rangle /2\geqslant 1\text{,}
\\
I(t)_{\text{A2}} &=&\langle \delta ^{2}(\hat{X}(t)_{\text{A1}}-\hat{X}(t)_{%
\text{A3}})\rangle /2+\langle \delta ^{2}(\hat{P}(t)_{\text{A1}}+g_{\text{A2}%
}\hat{P}(t)_{\text{A2}}+\hat{P}(t)_{\text{A3}})\rangle /2\geqslant 1\text{,}
\notag \\
I(t)_{\text{A3}} &=&\langle \delta ^{2}(\hat{X}(t)_{\text{A1}}-\hat{X}(t)_{%
\text{A2}})\rangle /2+\langle \delta ^{2}(\hat{P}(t)_{\text{A1}}+\hat{P}(t)_{%
\text{A2}}+g_{\text{A3}}\hat{P}(t)_{\text{A3}})\rangle /2\geqslant 1\text{.}
\notag
\end{eqnarray}
\end{widetext}

When any two in three inequalities are violated, the three atomic
ensembles are in an entangled state of GHZ-like type, where
$g_{\text{A1}}$, $g_{\text{A2}}$ and $g_{\text{A3}}$ are gain
factors for atomic ensembles.

If the mapping efficiencies $\eta _{_{\text{M}}}$ for three atomic memories
are the same, the gain factors in Eq. (10) should be the same, i.e. $g_{%
\text{A1}}=g_{\text{A2}}=g_{\text{A3}}=g_{\text{A}}$, and the values of
three inequalities are also identical, i.e. $I(t)_{\text{A1}}=I(t)_{\text{A2}%
}=I(t)_{\text{A3}}=I(t)_{\text{A}}$. According to Eqs. (9) and (10),
the combination of normalized quantum correlation variances for
atomic spin waves after a storage time $t$ is obtained:

\begin{widetext}
\begin{equation}
I(t)_{\text{A}}=\frac{12\text{e}^{-2r}+2(g_{\text{A}}+2)^{2}\text{e}%
^{-2r}+4(g_{\text{A}}-1)^{2}\text{e}^{2r}}{24}\eta _{_{\text{M}}}+(1+\frac{1%
}{4}g_{\text{A}})(1-\eta _{_{\text{M}}})\text{.}
\end{equation}
\end{widetext}

Similarly, by minimizing $I(t)_{\text{A}}$ we get the optimal
gain factor $g_{\text{A}}^{\text{opt}}$:

\begin{equation}
g_{\text{A}}^{\text{opt}}=\frac{2\eta _{_{\text{M}}}\text{e}^{4r}-2\eta _{_{%
\text{M}}}}{3\text{e}^{2r}+\eta _{_{\text{M}}}-3\eta _{_{\text{M}}}\text{e}%
^{2r}+2\text{e}^{4r}\eta _{_{\text{M}}}}\text{.}
\end{equation}

The smaller $I(t)_{\text{A}}$ is, the better the atomic entanglement
is. Fig. 4 shows the dependence of correlation variance combination
for three atomic ensembles on the squeezing parameter $r$ of initial
squeezed states and the mapping efficiency $\eta _{_{\text{M}}}$
with the storage time of 1000 ns. We can see that the combination is
reduced with the increase of squeezing parameter $r$ and mapping
efficiency $\eta _{_{\text{M}}}$. In our experiment, the squeezing
parameter $r$ and the mapping efficiency $\eta _{_{\text{M}}}$ are
0.38 and 23\%, respectively.

\begin{figure}[tbp]
\begin{center}
\includegraphics[width=8.6cm]{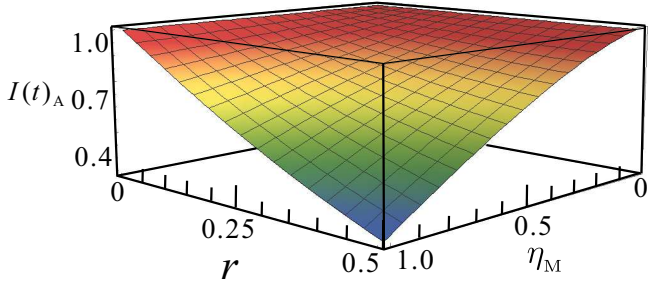}
\end{center}
\caption{The dependence of combination of normalized quantum correlation
variances among three atomic ensembles after a storage time of 1000 ns on the squeezing parameter $r$ of three DOPAs and the mapping
efficiency $\protect\eta _{_{\text{M}}}$, where the gains are taken as the
optimal gain $g_{\text{A1-A3}}^{\text{opt}}$.}
\end{figure}

\textbf{Quantum state transfer from stored entangled state of atomic spin
waves to released optical submodes.} The control optical beams of read
process enable to map quantum state from the atomic spin waves to
released optical submodes. Using Eqs. (2) and (9), the amplitude (phase)
quadratures $\hat{X}(\hat{P})(t)_{\text{L}j}$ of released optical submodes
after a storage time $t$ are calculated. We have

\begin{widetext}
\begin{eqnarray}
\hat{X}(t)_{\text{L}1} &=&\sqrt{\frac{\eta }{3}}\text{e}^{r}\hat{X}_{\text{S}%
1}^{(0)}+\sqrt{\frac{2\eta }{3}}\text{e}^{-r}\hat{X}_{\text{S}2}^{(0)}+\sqrt{%
1-\eta }\hat{X}_{\text{L}1}^{\text{vac}}\text{,} \\
\hat{P}(t)_{\text{L}1} &=&\sqrt{\frac{\eta }{3}}\text{e}^{-r}\hat{P}_{\text{S%
}1}^{(0)}+\sqrt{\frac{2\eta }{3}}\text{e}^{r}\hat{P}_{\text{S}2}^{(0)}+\sqrt{%
1-\eta }\hat{P}_{\text{L}1}^{\text{vac}}\text{,}  \notag \\
\hat{X}(t)_{\text{L}2} &=&\sqrt{\frac{\eta }{3}}\text{e}^{r}\hat{X}_{\text{S}%
1}^{(0)}-\sqrt{\frac{\eta }{6}}\text{e}^{-r}\hat{X}_{\text{S}2}^{(0)}+\sqrt{%
\frac{\eta }{2}}\text{e}^{-r}\hat{X}_{\text{S}3}^{(0)}+\sqrt{1-\eta }\hat{X}%
_{\text{L}2}^{\text{vac}}\text{,}  \notag \\
\hat{P}(t)_{\text{L}2} &=&\sqrt{\frac{\eta }{3}}\text{e}^{-r}\hat{P}_{\text{S%
}1}^{(0)}-\sqrt{\frac{\eta }{6}}\text{e}^{r}\hat{P}_{\text{S}2}^{(0)}+\sqrt{%
\frac{\eta }{2}}\text{e}^{r}\hat{P}_{\text{S}3}^{(0)}+\sqrt{1-\eta }\hat{P}_{%
\text{L}2}^{\text{vac}}\text{,}  \notag \\
\hat{X}(t)_{\text{L}3} &=&\sqrt{\frac{\eta }{3}}\text{e}^{r}\hat{X}_{\text{S}%
1}^{(0)}-\sqrt{\frac{\eta }{6}}\text{e}^{-r}\hat{X}_{\text{S}2}^{(0)}-\sqrt{%
\frac{\eta }{2}}\text{e}^{-r}\hat{X}_{\text{S}3}^{(0)}+\sqrt{1-\eta }\hat{X}%
_{\text{L}3}^{\text{vac}}\text{,}  \notag \\
\hat{P}(t)_{\text{L}3} &=&\sqrt{\frac{\eta }{3}}\text{e}^{-r}\hat{P}_{\text{S%
}1}^{(0)}-\sqrt{\frac{\eta }{6}}\text{e}^{r}\hat{P}_{\text{S}2}^{(0)}-\sqrt{%
\frac{\eta }{2}}\text{e}^{r}\hat{P}_{\text{S}3}^{(0)}+\sqrt{1-\eta }\hat{P}_{%
\text{L}3}^{\text{vac}}\text{.}  \notag
\end{eqnarray}

In this case of $g_{L1}^{\prime }=g_{L2}^{\prime }=g_{L3}^{\prime
}=g_{L}^{\prime }$, $I(t)_{L}$
in Eq. (3) can be calculated from Eq. (13):

\begin{equation}
I(t)_{\text{L}}=\frac{12\text{e}^{-2r}+2(g_{\text{L}}^{\prime }+2)^{2}\text{e%
}^{-2r}+4(g_{\text{L}}^{\prime }-1)^{2}\text{e}^{2r}}{24}\eta +(1+\frac{1}{4}%
g_{\text{L}}^{\prime })(1-\eta )\text{.}
\end{equation}
\end{widetext}

The optimal gain factor ($g_{\text{L}}^{\prime \text{opt}}$) for read out
process equals to:

\begin{equation}
g_{\text{L}}^{\prime \text{opt}}=\frac{2\eta \text{e}^{4r}-2\eta _{_{\text{M}%
}}^{\prime }}{3\text{e}^{2r}+\eta -3\eta \text{e}^{2r}+2\text{e}^{4r}\eta }%
\text{.}
\end{equation}

Fig. 2 is obtained from Eq. (14), where the optimal gain in Eq. (15)
is applied.

\textbf{Atomic ensemble.} The atomic energy levels of Rubidium used
for quantum memory medium is illustrated in Fig. 3b. The collective
coherence of the ground state $\left\vert 5S_{1/2},F=1\right\rangle $ and
meta state $\left\vert 5S_{1/2},F=2\right\rangle $ is used to store
nonclassical state of light. For balancing the storage efficiency and excess
noises from atoms, the frequencies of both signal and control optical beams
are detuned by $\Delta _{\text{s}}=700$ MHz (the detuning of signal optical
beam from the transition between energy levels g and e) and $\Delta _{\text{c%
}}=700.5$ MHz (the detuning of control optical beam from the transition
between energy levels m and e), respectively. The detuning is realized by
two sets of double-pass 1.7 GHz acousto-optical modulators (AOM$_{1-2}$).
Three rubidium vapour cells of 7.5 cm-long with 10 torr of neon
buffer gas in a $\mu $ metal magnetic shielding are used as atomic media,
where the neon buffer gas prevents thermal diffusion to increase the atomic
coherence. The three rubidium atomic cells are heated to around 65 \textbf{$%
^{\circ }$C} in our experiment.

\textbf{Experimental time sequence.} In the beginning of each period, the laser is turned on for 2 ms by
AOM$_{3}$ and split into three optical beams $\hat{a}_{\text{S1}}^{(0)}$, $\hat{a}_{\text{S2}%
}^{(0)}$ and $\hat{a}_{\text{S3}}^{(0)} $, which are used for input signals of
DOPAs respectively for persistently locking the length of optical cavities. Then the input signals of DOPAs are turned off for
20000 ns by AOMs$_{3}$ to generate three squeezed vacuum states of light. Within the locking
period the three entangled optical submodes are chopped into 500 ns pulses
with AOMs$_{4-6}$, which are used for input submodes $\hat{a}(0)_{\text{S1}}$, $\hat{a}(0)_{\text{S2}}$ and $\hat{a}(0)_{%
\text{S3}}$ of three atomic ensembles. Once the signal pulses enter into the
atomic cell, the control optical beams $\hat{a}_{\text{C1}}$, $\hat{a}_{\text{C2}}$
and $\hat{a}_{\text{C3}}$ are switched off by AOM$_{7}$ to complete the
quantum storage. At an user-controlled moment (1000 ns for our experiment) within storage lifetime, the control optical beams $\hat{a}%
_{\text{C1}}$, $\hat{a}_{\text{C2}}$ and $\hat{a}_{\text{C3}}$ are turned on
again by AOM$_{7}$ at to
obtain three released optical pulses $\hat{a}(t)_{\text{S1}}$, $\hat{a}(t)_{%
\text{S2}}$ and $\hat{a}(t)_{\text{S3}}$.

\section*{Data Availability}

The data that support the findings of this study are available from
the corresponding author on request.

\textbf{Acknowledgments}

The authors acknowledge financial support from the Key Project of the
Ministry of Science and Technology of China (Grant No. 2016YFA0301402), the
Natural Science Foundation of China (Grants Nos. 11322440, 11474190,
11304190, 11654002), FOK YING TUNG Education Foundation, the Program for
Sanjin Scholars of Shanxi Province.

\textbf{Author contributions}

X.J., H.W. and C.X. conceived the original idea. Z.Y., L.W., X.J. and K.P.
designed the experiment. Z.Y., L.W., Y.L. and R.D. constructed and performed
the experiment. Z.Y., X.J. and S.L. accomplished theoretical calculation and
the data analysis. Z.Y., X.J., C.X. and K.P. wrote the paper. All the
authors reviewed the manuscript.

\textbf{Additional information}

\textbf{Competing financial interests}: The authors declare no competing financial interests.

\end{document}